\documentclass[12pt]{article}
\usepackage[ansinew]{inputenc}
\usepackage{array}
\usepackage{color}
\usepackage{amsmath}
\usepackage{amsxtra}
\usepackage{amstext}
\usepackage{amssymb}
\usepackage{latexsym}
\begin{document}

\sloppy
\newcommand{\proof}{{\it Proof~}}
\newtheorem{thm}{Theorem}[section]
\newtheorem{cor}[thm]{Corollary}
\newtheorem{lem}[thm]{Lemma}
\newtheorem{prop}[thm]{Proposition}
\newtheorem{eg}[thm]{Example}
\newtheorem{defn}[thm]{Definition}

\newtheorem{rem}[thm]{Remark}
\numberwithin{equation}{section}

\thispagestyle{empty}
\begin{center}
{\bf\Large\textbf{On skew cyclic codes over $F_q+vF_q+v^2F_q$}}\\
\vspace{.3cm}
{Mohammad Ashraf and Ghulam Mohammad}\\
\vspace{.2cm}
 Department of Mathematics\\
    Aligarh Muslim University \\
        Aligarh -202002(India)\\
{\it E-mails} :  {mashraf80@hotmail.com; mohdghulam202@gmail.com}
\end{center}
\parindent=0mm
\begin{abstract}
\noindent In the present paper, we study skew cyclic codes over the ring $F_{q}+vF_{q}+v^2F_{q}$, where $v^3=v,~q=p^m$ and $p$ is an odd prime. We investigate the structural properties of skew cyclic codes over $F_{q}+vF_{q}+v^2F_{q}$ using decomposition method. By defining a Gray map from $F_{q}+vF_{q}+v^2F_{q}$ to $F_{q}^3$, it has been proved that the Gray image of a skew cyclic code of length $n$ over $F_{q}+vF_{q}+v^2F_{q}$ is a skew $3$-quasi cyclic code of length $3n$ over $F_{q}$. Further, it is shown that the skew cyclic codes over $F_{q}+vF_{q}+v^2F_{q}$ are principally generated. Finally, the idempotent generators of skew cyclic codes over $F_{q}+vF_{q}+v^2F_{q}$ are also obtained.
\end{abstract}

\vspace{.4cm}
\parindent=7mm

\section{Introduction} During the last decades of the twentieth century a great deal of attention has been given to the study of linear codes over finite rings because of their new role in algebraic coding theory and their successful applications. The class of cyclic codes is a very important class of linear codes from both theoretical and practical point of view which are easier to implement due to their rich algebraic structure. Cyclic codes have been studied for the last six decades. Based on these facts, cyclic codes have become one of the most important class in coding theory. A landmark paper by Hammons, et al. \cite{13} discovered that some good nonlinear codes over $\mathbb{Z}_2$ can be viewed as binary images under a Gray map of linear cyclic codes over $\mathbb{Z}_4$. But all this work is restricted to codes that are defined in a commutative ring.\\

Boucher et al. \cite{31}, \cite{32} and \cite{33} studied the structure of skew cyclic codes over a non commutative ring $F[x, \theta]$, called skew polynomial ring, where $F$ is a finite field and $\theta$ is a field automorphism of $F$. They generalized the class of linear and cyclic codes to the class of skew cyclic codes by using the ring $F[x, \theta]$, where the generator polynomials of skew cyclic codes come from the ring $F[x, \theta]$. They also gave some examples of skew cyclic codes with Hamming distances larger than the best known linear codes with the same parameters. Later on, Abualrub et al. \cite{27} and  Bhaintwal \cite{30}, defined skew quasi cyclic codes over these classes of rings. The main motivation of studying codes in this setting is that polynomials in skew polynomial rings exhibit many factorizations and hence there are many ideals in skew polynomial ring than in the commutative ring. But all this work is restricted to the condition that the order of the automorphism must be a factor of the length of the code. In \cite{38}, Siap, et al. removed this condition and they studied the structural properties of skew cyclic codes of arbitrary length over finite fields. A lot of work has been done in this direction (see references \cite{28,29,34}).\\

Recently, Jitman et al. \cite{37} defined skew constacyclic codes by defining the skew polynomial ring with coefficients from finite chain rings, especially the ring $F_{p^{m}}+uF_{p^{m}}$ where $u^2=0$. Further Gursoy et al. \cite{36} investigated the structural properties of skew cyclic codes through the decomposition method over $F_q+vF_q$, where $v^2=v$ and $q=p^m$. Very recently, the authors \cite{29} studied the structural properties of skew cyclic codes over the ring $F_3+vF_3$ with $v^2=1$ by considering the automorphism as; $\theta~:~v~\mapsto -v$. They proved that skew cyclic codes over $F_3+vF_3$ are equivalent to either cyclic codes or quasi cyclic codes. In the present paper, we study skew cyclic codes over the ring $F_{q}+vF_{q}+v^2F_{q}$, where $v^3=v,~q=p^m$ and $p$ is an odd prime by using the same technique as used by Gursoy et al. \cite{36} for the ring $F_q+vF_q$, where $v^2=v$ and $q=p^m$.\\

Throughout the paper $R$ will denote the ring $F_{q}+vF_{q}+v^2F_{q}$ with $v^3=v,~q=p^m$ and $p$ is an odd prime. Consider the automorphism $\theta_t:R\longrightarrow R$ such that $\theta_t(a+vb+v^2c)=a^{p^t}+vb^{p^t}+v^2c^{p^t}$. It is to be noted that $\theta_1$ is the Frobenius automorphism of $F_{q}$ and $\theta_t=\theta_{1}^{t}$. In this paper, we will use the automorphism $\theta_t$ instead of the automorphism $v\mapsto 1-v$ which was used by Gao in \cite{34}.

\vspace{.4cm}
\parindent=7mm

\section{Preliminaries} Let $R=F_{q}+vF_{q}+v^2F_{q},$ where $q=p^m$ and $p$ is an odd prime. $R$ is a commutative and non-chain ring with characteristic $p$ which contains $q^3$ elements. The ring is endowed with the natural addition and multiplication with the property $v^3=v$ and it can be viewed as the quotient ring $F_q[v]/\langle v^3-v\rangle$. The elements of $R$ can be uniquely written as $a+vb+v^2c,$ where $a,~b,~c\in F_{q}$. It is a semi-local ring having three maximal ideals $\langle v\rangle,~\langle v-1\rangle$ and $\langle v+1\rangle.$\\

Define a mapping $\theta_t:R\longrightarrow R$ such that $\theta_t(a+vb+v^2c)=a^{p^t}+vb^{p^t}+v^2c^{p^t},~\mbox{for all}~a,~b,~c\in F_{q}$. One can verify that $\theta_t$ is an automorphism on $R$ and $\theta_t=\theta_{1}^{t}$. This automorphism acts on $F_{q}$ as follows: $$\theta_t:F_{q}\longrightarrow F_{q}$$ $$a\mapsto a^{p^t}.$$ It may be noted that the order of this automorphism is $|\langle\theta_t\rangle|=m/t$ and the subring $F_{p^t}+vF_{p^t}+v^2F_{p^t}$ of $R$ is invariant under $\theta_t$.\\

\begin{defn} For a given automorphism $\theta_t$ of $R$, the set $R[x, \theta_t]=\{a_0+a_1x+a_2x^2+\cdots+a_nx^n|~a_i\in R, n\geq0\}$ of formal polynomials forms a ring under usual addition of polynomials and multiplication is defined by the rule $(ax^i)(bx^j)=a\theta_{t}^{i}(b)x^{i+j}$. The ring $R[x, \theta_t]$ is called skew polynomial ring over $R$.
\end{defn}

It can be easily seen that the ring $R[x, \theta_t]$ is non-commutative unless $\theta_t$ is the identity automorphism on $R$. Therefore, when an ideal of $R[x, \theta_t]$ is considered, one should specify whether it is a right ideal or a left ideal. The skew polynomial ring $R[x, \theta_t]$ is not left or right Euclidean. However, the division algorithm holds for some polynomials whose leading coefficients are invertible (for detail see references \cite{32} and \cite{37}).

\vspace{.4cm}
\parindent=7mm

\section{Gray map and linear codes over $R$} Gao \cite{35}, studied linear codes over the ring $F_p+uF_p+u^2F_p,$ where $u^3=u$ and $p$ is an odd prime. Here, we generalize his study to linear codes over the ring $R$. Let $R^n$ be the set of all $n$-tuples over $R$, then a nonempty subset $C$ of $R^n$ is called a code of length $n$ over $R$. $C$ is called linear code of length $n$ over $R$ if it is an $R$-submodule of $R^n$. Elements of $C$ are called codewords and therefore each codeword $c$ in such a code $C$ is just an $n$-tuple of the form $x=(x_0, x_1, \cdots, x_{n-1})\in R^n.$\\

The Hamming weight $w_H(x)$ of a codeword $x=(x_0, x_1, \cdots, x_{n-1})\in R^n$ is the number of nonzero components. The minimum weight $w_H(C)$ of a code $C$ is the smallest weight among all its nonzero codewords. For $x=(x_0, x_1, \cdots, x_{n-1}),~y=(y_0, y_1, \cdots, y_{n-1})\in R^n$,\\
$d_H(x, y)=|\{i~|~x_i\neq y_i\}|$ is called the Hamming distance between $x$ and $y\in R^n$ and is denoted by $$d_H(x, y)=w_H(x-y).$$
The minimum Hamming distance between distinct pairs of codewords of a code $C$ is called the minimum distance of $C$ and is denoted by $d_H(C)$ or shortly $d_H$.\\

Now, we define the Lee weight of an element $r=a+vb+v^2c\in R$ as follows: $$w_L(r)=w_H(a, a+b+c, a-b+c),$$ where $w_H$ denotes the usual Hamming weight on $F_q.$ Let $x=(x_0, x_1, \cdots, x_{n-1})$ be a vector in $R^n.$ Then the Lee weight of $x$ is the rational sum of Lee weights of its components, that is, $w_L(x)=\sum\limits_{i=0}^{n-1}w_L(x_i).$ For any elements $x, y\in R^n,$ the Lee distance is given by $d_L(x, y)=w_L(x-y).$ The minimum Lee distance of a code $C$ is the smallest nonzero Lee distance between all pairs of distinct codewords. The minimum Lee weight of $C$ is the smallest nonzero Lee weight among all codewords. If $C$ is linear, then the minimum Lee distance is the same as the minimum Lee weight.\\

The Gray map $\phi$ from $R$ to $F_{q}^3$ is defined as $\phi(a+vb+v^2c)=(a, a+b+c, a-b+c)$. It can be easily seen that $\phi$ is linear. The Gray map $\phi$ can be extended to $R^n$ in a natural way, that is, $\phi:R^n\longrightarrow F_{q}^{3n}$ such that $\phi(x_0, x_1, \cdots, x_{n-1})=(a_0, a_0+b_0+c_0, a_0-b_0+c_0, \cdots, a_{n-1}, a_{n-1}+b_{n-1}+c_{n-1}, a_{n-1}-b_{n-1}+c_{n-1})$, where $x_i=a_i+vb_i+v^2c_i$ for $i=0, 1, \cdots, n-1$.\\

\noindent The following property is obvious from the definition of the Gray map:

\begin{prop} The Gray map $\phi$ is a distance-preserving map or isometry from $R^n$(Lee distance) to $F_{q}^{3n}$(Hamming distance) and it is also $F_{q}$-linear.
\end{prop}

For a code $C$ over $R,$ define $$C_1=\{a\in F_{q}^n~|~a+vb+v^2c\in C~\mbox{some}~b, c\in F_{q}^n \},$$
$$C_2=\{a+b+c\in F_{q}^n~|~a+vb+v^2c\in C \},$$ and $$C_3=\{a-b+c\in F_{q}^n~|~a+vb+v^2c\in C \}.$$
If $C$ is linear code of length $n$ over $R$, then $C_1,~C_2$ and $C_3$ are all linear codes of length $n$ over $F_q.$ Moreover, the linear code $C$ of length $n$ over $R$ can be uniquely expressed as $$C=(1-v^2)C_1\oplus\frac{p+1}{2}(v^2+v)C_2\oplus\frac{p+1}{2}(v^2-v)C_3.$$\\

A generator matrix of $C$ is a matrix whose rows generate $C$. Let $$C=(1-v^2)C_1\oplus\frac{p+1}{2}(v^2+v)C_2\oplus\frac{p+1}{2}(v^2-v)C_3$$ be a linear code of length $n$ over $R$ with generator matrix $G.$ Then $G$ can be written as
$$G=\left(
\begin{array}{ccccc}
(1-v^2)G_1 \\
\\
\frac{p+1}{2}(v^2+v)G_2\\
\\
\frac{p+1}{2}(v^2-v)G_3

\end{array}\right),$$
where $G_1,~G_2$ and $G_3$ are the generator matrices of $C_1,~C_2$ and $C_3$ respectively.\\

Let $x=(x_0, x_1, \cdots, x_{n-1})$ and $y=(y_0, y_1, \cdots, y_{n-1})$ be two elements of $R^n$. Then the Euclidean inner product of $x$ and $y$ in $R^n$ is defined as $$x\cdot y=x_0y_0+x_1y_1+\cdots+x_{n-1}y_{n-1}.$$
The dual code $C^\perp$ of $C$ is defined as $$C^\perp=\{x\in R^n|~x\cdot y=0,~\mbox{for~all}~y\in C\}.$$
A code $C$ is called self-orthogonal if $C\subseteq C^\perp$ and self dual if $C=C^\perp$.\\

\noindent Now, we give some results on linear codes over $R$, which are the generalization of results on linear codes over $F_p+vF_p+v^2F_p$. So, we are omitting the proofs of the results.

\begin{thm} If $C=(1-v^2)C_1\oplus\frac{p+1}{2}(v^2+v)C_2\oplus\frac{p+1}{2}(v^2-v)C_3$ is a linear code of length $n$ over $R$, then $\phi(C)=C_1\otimes C_2\otimes C_3$ and $|C|=|C_1||C_2||C_3|$.
\end{thm}

\begin{cor} Let $C=(1-v^2)C_1\oplus\frac{p+1}{2}(v^2+v)C_2\oplus\frac{p+1}{2}(v^2-v)C_3$ be a linear code of length $n$ over $R$, where $C_i$ is a linear code with dimension $k_i$ and minimum Hamming distance $d(C_i)$ for $i=1, 2, 3$. Then $\phi(C)$ is a linear code with parameters \linebreak$[3n, k_1+k_2+k_3,~min\{d(C_1), d(C_2), d(C_3)\}]$ over $F_{q}$.
\end{cor}

\noindent One of the properties of the Gray map we defined is that it preserves the duality as given in the following lemma:

\begin{lem} Let $C^\perp$ be the dual code of $C$ over $R$. Then $\phi(C^\perp)=\phi(C)^\perp$. In particular, if $C$ is self-dual, then so is $\phi(C)$.
\end{lem}
{\bf \it{Proof.}} Let $x_1=a_1+vb_1+v^2c_1$ and $x_2=a_2+vb_2+v^2c_2\in C$, where $a_1, b_1, c_1, a_2, b_2, c_2\in F_{q}^n$. Now by Euclidean inner product of $x_1$ and $x_2$, we have

\[\begin{split}
 x_1\cdot x_2 &=(a_1+vb_1+v^2c_1)\cdot(a_2+vb_2+v^2c_2)\\
       &=a_1a_2+v(a_1b_2+a_2b_1+b_1c_2+b_2c_1)+v^2(a_1c_2+a_2c_1+b_1b_2+c_1c_2).\\
\end{split}\]
Since $C$ is a self-dual code, $C= C^\perp$, we find that $a_1a_2=a_1b_2+a_2b_1+b_1c_2+b_2c_1=a_1c_2+a_2c_1+b_1b_2+c_1c_2=0$. Now
$$\phi(x_1)\phi(x_2)=(a_1, a_1+b_1+c_1, a_1-b_1+c_1)(a_2, a_2+b_2+c_2, a_2-b_2+c_2)=0.$$ Thus $\phi(C^\perp)\subseteq \phi(C)^\perp$. On the other hand let $|C|={(q)}^{k_1+k_2+k_3}$ and $C$ is of length $n$. Then $\phi(C)$ has the parameters $[3n, k_1+k_2+k_3]$. Since $|\phi(C)|=|C|$, $|\phi(C)^\perp|={(q)}^{3n-(k_1+k_2+k_3)}$. Further $|\phi(C^\perp)|=|C^\perp|=q^{3n}/|C|=q^{3n-(k_1+k_2+k_3)}$. Hence $\phi(C^\perp)=\phi(C)^\perp$.\\

\noindent In view of the previous lemma, the following theorem can be easily obtained:

\begin{thm} Let $C$ be a linear code of length $n$ over $R$ and let $\phi(C)=C_1\otimes C_2\otimes C_3$. Then $C$ can be uniquely expressed as $C=(1-v^2)C_1\oplus\frac{p+1}{2}(v^2+v)C_2\oplus\frac{p+1}{2}(v^2-v)C_3$. Furthermore, if $\phi(C^\perp)=C_1^\perp\otimes C_2^\perp\otimes C_3^\perp$, then $C^\perp=(1-v^2)C_1^\perp\oplus\frac{p+1}{2}(v^2+v)C_2^\perp\oplus\frac{p+1}{2}(v^2-v)C_3^\perp$.
\end{thm}

\vspace{.4cm}
\parindent=7mm

\section{Skew cyclic codes over $R$} In the present section, we study skew cyclic codes over $R$. Let $\theta_t$ be an automorphism on $R$ given by $\theta_t(a+vb+v^2c)=a^{p^t}+vb^{p^t}+v^2c^{p^t}$. Then a linear code $C$ of length $n$ over $R$ is called a skew cyclic code or $\theta_t$-cyclic code if it satisfies the property
$c=(c_0, c_1, \cdots, c_{n-1})\in C~\mbox{implies}~\sigma(c)=(\theta_t(c_{n-1}), \theta_t(c_0), \cdots, \theta_t(c_{n-2}))\in C$, where $\sigma(c)$ denotes the skew cyclic shift of $c$.\\

In \cite{38}, it was shown that a linear code $C$ of length $n$ over $F_{q}$ is a skew cyclic code with respect to automorphism $\theta$ if and only if it is a left $F_{q}[x, \theta]$-submodule of $F_{q}[x, \theta]/\langle x^n-1 \rangle$. Moreover, if $C$ is a left submodule of $F_{q}[x, \theta]/\langle x^n-1 \rangle$, then $C$ is generated by a monic polynomial $g(x)$ which is a right divisor of $x^n-1$ in $F_{q}[x, \theta]$.\\

The method which we use in this section is same as the method used by Gao in \cite{35} over the ring $F_p+vF_p+v^2F_p$ with $v^3=v$. The main difference in our case is that the ring $R[x, \theta_t]$ is non-commutative.

\begin{thm} Let $C=(1-v^2)C_1\oplus\frac{p+1}{2}(v^2+v)C_2\oplus\frac{p+1}{2}(v^2-v)C_3$ be a linear code of length $n$ over $R$. Then $C$ is a skew cyclic code over $R$ with respect to automorphism $\theta_t$ if and only if $C_{1}, C_{2}$ and $C_{3}$ are skew cyclic codes of length $n$ over $F_{q}$ with respect to same automorphism $\theta_t$.
\end{thm}
{\bf \it{Proof.}} For any $r=(r_0, r_1, \cdots, r_{n-1})\in C$, we can write its components as $r_i=(1-v^2)a_i+\frac{p+1}{2}(v^2+v)b_i+\frac{p+1}{2}(v^2-v)c_i$, where $a_i,~b_i$, $c_i\in F_{q},~0\leq i\leq n-1$. Let $a=(a_0, a_1, \cdots, a_{n-1}),~b=(b_0, b_1, \cdots, b_{n-1})$ and $c=(c_0, c_1, \cdots, c_{n-1})$. Then $a\in C_{1},~b\in C_2$ and $c\in C_{3}$. Now, Suppose $C_{1},~C_2$ and $C_{3}$ are skew cyclic codes over $F_{q}$ with respect to automorphism $\theta_t$. This means that $\sigma(a)=(\theta_t(a_{n-1}), \theta_t(a_0), \cdots, \theta_t(a_{n-2}))=(a_{n-1}^{p^t}, a_0^{p^t}, \cdots, a_{n-2}^{p^t})\in C_{1},~ \sigma(b)=(\theta_t(b_{n-1}), \theta_t(b_0), \cdots, \theta_t(b_{n-2}))=(b_{n-1}^{p^t}, b_0^{p^t}, \cdots, b_{n-2}^{p^t})\in C_{2}$ and $\sigma(c)=(\theta_t(c_{n-1}), \theta_t(c_0), \cdots, \theta_t(c_{n-2}))=(c_{n-1}^{p^t}, c_0^{p^t}, \cdots, c_{n-2}^{p^t})\in C_{3}$. Thus $(1-v^2)\sigma(a)+(v^2+v)\frac{p+1}{2}\sigma(b)+(v^2-v)\frac{p+1}{2}\sigma(c)\in C$. It can be easily seen that $(1-v^2)\sigma(a)+(v^2+v)\frac{p+1}{2}\sigma(b)+(v^2-v)\frac{p+1}{2}\sigma(c)=\sigma(r)$. Hence $\sigma(r)\in C$, which means that $C$ is a skew cyclic code over $R$ with respect to automorphism $\theta_t$.\\

Conversely, suppose $C$ is a skew cyclic code over $R$ with respect to automorphism $\theta_{t}$. Let $r_i=(1-v^2)a_i+\frac{p+1}{2}(v^2+v)b_i+\frac{p+1}{2}(v^2-v)c_i$, for any $a=(a_0, a_1, \cdots, a_{n-1})\in C_{1},~b=(b_0, b_1, \cdots, b_{n-1})\in C_{2}$ and $c=(c_0, c_1, \cdots, c_{n-1})\in C_3$. Then $r=(r_0, r_1, ..., r_{n-1})\in C$. By the hypothesis $\sigma(r)\in C$. Since $(1-v^2)\sigma(a)+(v^2+v)\frac{p+1}{2}\sigma(b)+(v^2-v)\frac{p+1}{2}\sigma(c)=\sigma(r)$, $(1-v^2)\sigma(a)+(v^2+v)\frac{p+1}{2}\sigma(b)+(v^2-v)\frac{p+1}{2}\sigma(c)\in C$. Thus $\sigma(a)\in C_{1},~\sigma(b)\in C_{2}$ and $\sigma(c)\in C_3$, which implies that $C_{1},~C_2$ and $C_{3}$ are skew cyclic codes of length $n$ over $F_{q}$ with respect to automorphism $\theta_t$.

\begin{cor} Let $C$ be a skew cyclic code of length $n$ over $R$. Then the dual code $C^\perp$ is also a skew cyclic code of length $n$ over $R$.
\end{cor}
{\bf \it{Proof.}} In view of Theorem 3.5, we know that $C^\perp=(1-v^2)C_1^\perp\oplus\frac{p+1}{2}(v^2+v)C_2^\perp\oplus\frac{p+1}{2}(v^2-v)C_3^\perp$. Since the dual code of every skew cyclic code over $F_{q}$ is also skew cyclic (\cite{33}, Corollary 18), by Theorem 4.1, $C^\perp$ is a skew cyclic code over $R$.

\begin{cor} A code $C=(1-v^2)C_1\oplus\frac{p+1}{2}(v^2+v)C_2\oplus\frac{p+1}{2}(v^2-v)C_3$ of length $n$ over $R$ is a self-dual skew cyclic if and only if $C_1,~C_2$ and $C_3$ are self-dual skew cyclic codes of length $n$ over $F_{q}$.
\end{cor}

Let $C^\prime$ be a linear code of length $n$ over $F_{q}$ and $c=(c^1|c^2|\cdots |c^s)$ be a codeword in $C^\prime$ into $s$ equal parts of length $r$ where $n=rs$. If $(\sigma(c^1)|\sigma(c^2)|\cdots |\sigma(c^s))\in C^\prime$, then the linear code $C$ which is permutation equivalent to $C^\prime$ is called a skew quasi-cyclic code of index $s$ or skew $s$-quasi cyclic code. (for detail see reference \cite{27})

\begin{thm} Let $C$ be a skew cyclic code of length $n$ over $R$. Then $\phi(C)$ is a skew $3$-quasi cyclic code of length $3n$ over $F_{q}$.
\end{thm}
{\bf \it{Proof.}} In view of Theorem 3.2 and the definition of skew quasi-cyclic codes, we can obtain the required result.

\begin{thm} Let $C=(1-v^2)C_1\oplus\frac{p+1}{2}(v^2+v)C_2\oplus\frac{p+1}{2}(v^2-v)C_3$ be skew cyclic code of length $n$ over $R$. Then $C=\langle(1-v^2)g_1(x), \frac{p+1}{2}(v^2+v)g_2(x), \frac{p+1}{2}(v^2-v)g_3(x)\rangle$ and $|C|={q}^{3n-deg(g_1(x))-deg(g_2(x))-deg(g_3(x))}$, where $g_1(x),~g_2(x)$ and $g_3(x)$ are the generator polynomials of $C_{1},~C_2$ and $C_{3}$ respectively.
\end{thm}
{\bf \it{Proof.}} Since $C_{1}=\langle g_1(x)\rangle\subseteq{F_{q}[x, \theta_t]}/{\langle x^n-1\rangle},~C_{2}=\langle g_2(x)\rangle\subseteq{F_{q}[x, \theta_t]}/{\langle x^n-1\rangle},~C_3=\langle g_3(x)\rangle\subseteq{F_{q}[x, \theta_t]}/{\langle x^n-1\rangle}$ and $C=(1-v^2)C_1\oplus\frac{p+1}{2}(v^2+v)C_2\oplus\frac{p+1}{2}(v^2-v)C_3$, we find that $C=\{c(x)~|~c(x)=(1-v^2)f_1(x)+\frac{p+1}{2}(v^2+v)f_2(x)+\frac{p+1}{2}(v^2-v),~f_1(x)\in C_{1},~f_2(x)\in C_{2},~f_3(x)\in C_3\}.$ Therefore $$C\subseteq\langle(1-v^2)g_1(x), \frac{p+1}{2}(v^2+v)g_2(x), \frac{p+1}{2}(v^2-v)g_3(x)\rangle\subseteq R[x, \theta_t]/\langle x^n-1\rangle.$$
For any $(1-v^2)k_1(x)g_1(x)+\frac{p+1}{2}(v^2+v)k_2(x)g_2(x)+\frac{p+1}{2}(v^2-v)k_3(x)g_3(x)\in\linebreak\langle(1-v^2)g_1(x), \frac{p+1}{2}(v^2+v)g_2(x), \frac{p+1}{2}(v^2-v)g_3(x)\rangle\subseteq R[x, \theta_t]/\langle x^n-1\rangle,$ where $k_1(x), k_2(x), k_3(x)\in R[x, \theta_t]/\langle x^n-1\rangle$, there are $r_1(x), r_2(x), r_3(x)\in F_{q}[x, \theta_t]$ such that $$(1-v^2)k_1(x)=(1-v^2)r_1(x),$$ $$\frac{p+1}{2}(v^2+v)k_2(x)=\frac{p+1}{2}(v^2+v)r_2(x)$$ and $$\frac{p+1}{2}(v^2-v)k_3(x)=\frac{p+1}{2}(v^2-v)r_3(x).$$ This means that $$\langle(1-v^2)g_1(x), \frac{p+1}{2}(v^2+v)g_2(x), \frac{p+1}{2}(v^2-v)g_3(x)\rangle\subseteq C.$$ Hence $\langle(1-v^2)g_1(x), \frac{p+1}{2}(v^2+v)g_2(x), \frac{p+1}{2}(v^2-v)g_3(x)\rangle=C$. Since $|C|=|C_{1}||C_{2}||C_3|$, $|C|={q}^{3n-deg(g_1(x))-deg(g_2(x))-deg(g_3(x))}$.

\begin{thm} Let $C_1,~C_2$ and $C_3$ be skew cyclic codes over $F_{q}$ with monic generator polynomials $g_1(x),~g_2(x)$ and $g_3(x)$ respectively. If $C=(1-v^2)C_1\oplus\frac{p+1}{2}(v^2+v)C_2\oplus\frac{p+1}{2}(v^2-v)C_3$ is a skew cyclic code of length $n$ over $R$, then there is a unique polynomial $g(x)\in R[x, \theta_t]$ such that $C=\langle g(x)\rangle$ and $g(x)$ is a right divisor of $x^n-1$, where $g(x)=(1-v^2)g_1(x)+\frac{p+1}{2}(v^2+v)g_2(x)+ \frac{p+1}{2}(v^2-v)g_3(x)$.
\end{thm}
{\bf \it{Proof.}} By Theorem 4.5, we may assumed that $C=\langle(1-v^2)g_1(x), \frac{p+1}{2}(v^2+v)g_2(x), \frac{p+1}{2}(v^2-v)g_3(x)\rangle$, where $g_1(x),~g_2(x)$ and $g_3(x)$ are the monic generator polynomials of $C_{1},~C_2$ and $C_{3}$ respectively. Let $g(x)=(1-v^2)g_1(x)+\frac{p+1}{2}(v^2+v)g_2(x)+ \frac{p+1}{2}(v^2-v)g_3(x)$. Clearly, $\langle g(x)\rangle\subseteq C $. Note that $$(1-v^2)g_1(x)=(1-v^2)g(x),$$ $$\frac{p+1}{2}(v^2+v)g_2(x)=\frac{p+1}{2}(v^2+v)g(x)$$ and $$\frac{p+1}{2}(v^2-v)g_3(x)=\frac{p+1}{2}(v^2-v)g(x),$$ so $C\subseteq \langle g(x)\rangle$. Hence $C=\langle g(x)\rangle$. Since $g_1(x),~g_2(x)$ and $g_3(x)$ are monic right divisors of $x^n-1$, there are $r_1(x), r_2(x), r_3(x)\in F_{q}[x, \theta_t]/\langle x^n-1\rangle$ such that $$x^n-1=r_1(x)g_1(x)=r_2(x)g_2(x)=r_3(x)g_3(x).$$ This implies that $$x^n-1=[(1-v^2)r_1(x)+\frac{p+1}{2}(v^2+v)r_2(x)+\frac{p+1}{2}(v^2-v)r_3(x)]g(x).$$ Hence, $g(x)|x^n-1$. The uniqueness of $g(x)$ can be followed from that of $g_1(x), g_2(x)$ and $g_3(x)$.\\

\noindent The following corollary is an immediate consequence of the above theorem:

\begin{cor} Every left submodule of $R[x, \theta_t]/\langle x^n-1\rangle$ is principally generated.
\end{cor}

\noindent In order to study the generator polynomials of the dual code of a skew cyclic code over $R$, we need the following definition which can be found in \cite{33}.\\

\noindent Let $g(x)=g_0+g_1x+\cdots+g_rx^r$ and $h(x)=h_0+h_1x+\cdots+h_{n-r}x^{n-r}$ be polynomials in $F_{q}[x, \theta_t]$ such that $x^n-1=h(x)g(x)$ and $C^\prime$ be the skew cyclic code generated by $g(x)$ in $F_{q}[x, \theta_t]/\langle x^n-1\rangle$. Then the dual code of $C^\prime$ is a skew cyclic code generated by the polynomial $\bar{h}(x)=h_{n-r}+\theta_t(h_{n-r-1})x+\cdots+\theta_{t}^{n-r}(h_0)x^{n-r}$.\\

\noindent In view of Theorems 3.5 $\&$ 4.6, we have the following corollary:

\begin{cor} Let $C_1,~C_2$ and $C_3$ be skew cyclic codes over $F_{q}$ and $g_1(x),~g_2(x)$ and $g_3(x)$ be their generator polynomials such that $$x^n-1=h_1(x)g_1(x)=h_2(x)g_2(x)=h_3(x)g_3(x) \in F_{q}[x, \theta_t].$$ If $C=(1-v^2)C_1\oplus\frac{p+1}{2}(v^2+v)C_2\oplus\frac{p+1}{2}(v^2-v)C_3$, then $$C^\perp=\langle(1-v^2)\bar{h_1}(x)+\frac{p+1}{2}(v^2+v)\bar{h_2}(x)+\frac{p+1}{2}(v^2-v)\bar{h_3}(x)\rangle$$ and $|C^\perp|={q}^{deg(g_1(x))+deg(g_2(x))+deg(g_3(x))}$.
\end{cor}

\section{Idempotent generators of skew cyclic codes over $R$} The idempotent generators of skew cyclic codes over $F_q$ studied by Gursoy et al. \cite{36} under some restrictions. In fact, they proved the following results:

\begin{lem}{\cite[Lemma 2]{36}} Let $g(x)\in F_{q}[x, \theta_t]$ be a monic right divisor of $x^n-1$. If $g.c.d.(n, m_t)=1$, then $g(x)\in F_{p^t}[x]$, where $m_t=m/t$ denotes the order of the automorphism $\theta_t$.
\end{lem}

\begin{lem}{\cite[Theorem 6]{36}} Let $g(x)\in F_{q}[x, \theta_t]$ be a monic right divisor of $x^n-1$ and $C=\langle g(x)\rangle$. If $g.c.d.(n, m_t)=1$ and $g.c.d.(n, q)=1$, then there exists an idempotent polynomial $e(x)\in F_{q}[x, \theta_t]/\langle x^n-1\rangle$ such that $C=\langle e(x)\rangle$.
\end{lem}

\noindent Now, we give the idempotent generators of skew cyclic codes over $R$.

\begin{thm} Let $C=(1-v^2)C_1\oplus\frac{p+1}{2}(v^2+v)C_2\oplus\frac{p+1}{2}(v^2-v)C_3$ be skew cyclic code of length $n$ over $R$ and $g.c.d.(n, m_t)=1,~g.c.d.(n, q)=1$. Then $C_i$ has idempotent generator, say $e_i(x)$ for $i=1, 2, 3$. Moreover $e(x)=(1-v^2)e_1(x)+\frac{p+1}{2}(v^2+v)e_2(x)+\frac{p+1}{2}(v^2-v)e_3(x)$ is an idempotent generator of $C$, that is, $C=\langle e(x)\rangle$.
\end{thm}
{\bf \it{Proof.}} In the light of Theorem 4.6 and Lemma 5.2 , the proof follows.\\

\noindent The following theorem gives the number of skew cyclic codes of length $n$ over $R$.

\begin{thm} Let $g.c.d.(n, m_t)=1$ and $x^n-1=\prod\limits_{i=1}^{r}{g_i^{s_i}(x)}$, where $g_i(x)\in F_{q}[x, \theta_t]$ is irreducible. Then the number of skew cyclic codes of length $n$ over $R$ is $\prod\limits_{i=1}^{r}{(s_i+1)^3}$.
\end{thm}
{\bf \it{Proof.}} In view of Lemma 5.1 if $g.c.d.(n, m_t)=1$, then $g_i(x)\in F_{p^t}[x]$. In this case the number of skew cyclic codes of length $n$ over $F_{q}$ is $\prod\limits_{i=1}^{r}(s_i+1)$. Since $C=(1-v^2)C_1\oplus\frac{p+1}{2}(v^2+v)C_2\oplus\frac{p+1}{2}(v^2-v)C_3$, $\prod\limits_{i=1}^{r}{(s_i+1)^3}$ is the number of skew cyclic codes of length $n$ over $R$. When $g.c.d.(n, m_t)\neq1$, the factorization of $x^n-1$ is not unique in $F_{q}[x, \theta_t]$, therefore we can not say anything certain about the number of skew cyclic codes in this case.\\

\noindent Now, we close our discussion with the following examples:\\

\noindent\textbf{Example 4.11} Let $R=F_9+vF_9$ be the ring with $v^3=v$ and $\theta$ be the Frobenius automorphism over $F_9$, that is, $\theta(r)=r^3$ for any $r\in F_9$, where $F_9=F_3[2\alpha+1],~\alpha^2=-1$. Then $$x^4-1=(x+1)(x+2)(x+\alpha)(x+2\alpha)\in F_9[x, \theta].$$ If $g_1(x)=g_2(x)=g_3(x)=x+2\alpha$, then $C_1=\langle g_1(x)\rangle,~C_2=\langle g_2(x)\rangle$ and $C_3=\langle g_3(x)\rangle$ are the skew cyclic codes over $F_9$ with parameters $[4, 3, 2]$. Therefore, the code $C=\langle (1-v^2)g_1(x)+\frac{p+1}{2}(v^2+v)g_2(x)+\frac{p+1}{2}(v^2-v)g_3(x)\rangle=\langle x+2\alpha\rangle$ is a skew cyclic code of length $4$ over $R$. Further, the Gray image $\phi(C)$ of $C$ is a skew 3-quasi cyclic code over $F_9$ with parameters $[12, 9, 2]$, which is an optimal code.\\

\noindent\textbf{Example 4.12} Let $R=F_9+vF_9$ be the ring with $v^3=v$ and $\theta$ be the Frobenius automorphism over $F_9$, that is, $\theta(r)=r^3$ for any $r\in F_9$, where $F_9=F_3[2\alpha+1],~\alpha^2=-1$. Then $$x^5-1=(x+2)(x^4+x^3+x^2+x+1)\in F_9[x, \theta].$$ Since $g.c.d.(5, 2)=1$, there exist $63$ nonzero skew cyclic codes of length $5$ over $R$.\\

Let $g_1(x)=g_2(x)=g_3(x)=x+2$. Then $C_1=\langle g_1(x)\rangle,~C_2=\langle g_2(x)\rangle$ and $C_3=\langle g_3(x)\rangle$ are the skew cyclic codes of length $5$ over $F_9$. Therefore, the code $C=\langle(1-v^2)g_1(x)+\frac{p+1}{2}(v^2+v)g_2(x)+\frac{p+1}{2}(v^2-v)g_3(x)\rangle=\langle x+2\rangle$ is a skew cyclic code of length $5$ over $R$. Also, the Gray image $\phi(C)$ of $C$ is a skew 3-quasi cyclic code of length $15$ over $F_9$.\\

\noindent\textbf{Example 4.13} Let $R=F_9+vF_9$ be the ring with $v^3=v$ and $\theta$ be the Frobenius automorphism over $F_9$, that is, $\theta(r)=r^3$ for any $r\in F_9$, where $F_9=F_3[2\alpha+1],~\alpha^2=-1$. Then

\[\begin{split}
 x^6-1&=(2+\alpha x+2\alpha x^3+x^4)(1+\alpha x+x^2)\\
       &=(2+x+(1+2\alpha)x^2+x^3)(1+x+(2\alpha+2)x^2+x^3)\\
       &\in F_9[x, \theta].
\end{split}\]
If $g_1(x)=g_2(x)=2+\alpha x+2\alpha x^3+x^4$ and $g_3(x)=2+x+(1+2\alpha)x^2+x^3$, then $C_1=\langle g_1(x)\rangle,~C_2=\langle g_2(x)\rangle$ and $C_3=\langle g_3(x)\rangle$ are the skew cyclic codes of length $6$ over $F_9$ with dimensions $2,~2$ and $3$ respectively. Thus the code $$C=\langle(1-v^2)g_1(x)+\frac{p+1}{2}(v^2+v)g_2(x)+\frac{p+1}{2}(v^2-v)g_3(x)\rangle$$ is a skew cyclic code of length $6$ over $R$. Also, the Gray image $\phi(C)$ of $C$ is a skew 3-quasi cyclic code over $F_9$ with parameters $[18, 7, 4]$.

\section{Conclusion} In this paper, we have studied the structural properties of skew cyclic codes over the ring $F_{q}+vF_{q}+v^2F_{q}$ by taking the automorphism $\theta_t:a+vb+v^2c\mapsto a^{p^t}+vb^{p^t}+v^2c^{p^t}$. We have proved that the Gray image of a skew cyclic code of length $n$ over $F_{q}+vF_{q}+v^2F_{q}$ is a skew $3$-quasi cyclic code of length $3n$ over $F_{q}$. It has also been shown that skew cyclic codes over $F_{q}+vF_{q}+v^2F_{q}$ are principally generated. Further, we have obtained idempotent generators of skew cyclic codes over $F_{q}+vF_{q}+v^2F_{q}$.

\begin{center}

\end{center}

\end{document}